\documentstyle[twocolumn,aps]{revtex}

\begin{document}
\draft
\title{Dependence of the vortex configuration on the geometry of mesoscopic flat
samples}
\author{B. J. Baelus and F. M. Peeters\cite{peeters}}
\address{Departement Natuurkunde, Universiteit Antwerpen (UIA), Universiteitsplein 1,%
\\
B-2610 Antwerpen, Belgium}
\date{\today}
\maketitle

\begin{abstract}
The influence of the geometry of a thin superconducting sample on the
penetration of the magnetic field lines and the arrangement of vortices are
investigated theoretically. We compare superconducting disks, squares and
triangles with the same surface area having nonzero thickness. The coupled
nonlinear Ginzburg-Landau equations are solved self-consistently and the
important demagnetization effects are taken into account. We calculate and
compare quantities like the free energy, the magnetization, the Cooper-pair
density, the magnetic field distribution and the superconducting current
density for the three geometries. For given vorticity the vortex lattice is
different for the three geometries, i.e. it tries to adapt to the geometry
of the sample. This also influences the stability range of the different
vortex states. For certain magnetic field ranges we found a coexistence of a
giant vortex placed in the center and single vortices toward the corners of
the sample. Also the $H-T$ phase diagram is obtained.
\end{abstract}


\section{Introduction}

In mesoscopic samples there is a competition between a triangle
configuration of the vortex lattice as being the lowest energy configuration
in bulk material (and films) and the boundary which tries to impose its
geometry on the vortex lattice. For example a circular geometry will favour
vortices situated on a ring near the boundary and only far away from the
boundary its influence diminishes and the triangular lattice may reappear.
Therefore, it is expected that different geometries will favour different
arrangements of vortices and will make certain vortex configurations more
stable than others. In small systems vortices may overlap so strongly that
it is more favourable to form one big giant vortex. The latter will
preferably have a circular geometry. As a consequence it is expected that
the giant to multivortex transition will be strongly influenced by the
geometry of the boundary as will be also the stability of the giant vortex
configuration.

These issues, the dependence of the stability of the giant vortex
configuration and of the different multivortex configurations on the
geometry of the sample will be investigated in the present paper. As an
example, we will compare the most important geometries: the circular disk,
the square and the triangle.

Mesoscopic (circular) disks and rings have been the most popular in this
respect. Experimentally, the magnetization of superconducting disks has been
measured as a function of the externally applied magnetic field \cite
{Buisson,Geim}. Several transitions between different superconducting states
were found and the magnetization depends sensitively on size and
temperature. The main part of the theoretical studies covered disks \cite
{Benoist,Palacios,PRL83,PRB62,Akkermans} and rings \cite
{Berger,Bruyndoncx1,Baelus} of zero thickness. In this case one can neglect
the magnetic field induced by the supercurrents and one assumes that the
total magnetic field equals the external applied magnetic field, which is
uniform. A limited number of studies considered disks \cite
{PRL79,PRB57,PRL81,PRB60} and rings \cite{PRB61} with {\it finite}
thickness. Then, the finite thickness of the disks influences the magnetic
field profile and it is necessary to take into account the demagnetization
effects. Often only the (circular symmetric) giant vortex states or the
superconducting/normal transition were investigated. Even in type-I
superconductors multivortex states in disks \cite{Palacios,PRL83,PRB62,PRL81}
and rings \cite{PRB61,Baelus} were predicted. It was found that if the disk
or the ring is large enough, the giant vortex nucleates into a multivortex
state in which the vortices are situated on a ring. In a ring geometry, we
found that breaking the circular symmetry through a non-central location of
the hole favors the multivortex state \cite{PRB61}. This means that by
changing the geometry, the giant vortex state transits into a multivortex
state.

Mesoscopic superconductors with non-circular geometries have attracted less
attention. Moshchalkov {\it et al} \cite{Moshchalkov} measured the
superconducting/normal transition in superconducting lines, squares and
square rings using resistance measurements. Bruyndoncx {\it et al} \cite
{Bruyndoncx2} calculated the $H-T$ phase diagram for a square with zero
thickness in the framework of the linearized Ginzburg-Landau theory, which
is only valid near the superconducting/normal boundary. They compared their
results with the $H-T$ phase boundary obtained from resistance measurements.
Fomin {\it et al} \cite{Fomin} studied square loops with leads attached to
it and found inhomogeneous Cooper-pair distributions in the loop with
enhancements near the corners of the square loop. Schweigert {\it et al} 
\cite{PRB57,PRB60} calculated the nucleation field as a function of the
sample area for disks, squares and triangles with zero thickness. Jadallah 
{\it et al} \cite{Jadallah} computed the superconducting/normal transition
for mesoscopic disks and squares of zero thickness. For macroscopic squares,
the magnetic field distribution and the flux penetration are investigated in
detail by experimental observations using the magneto-optical Faraday effect
and by first-principles calculations which describe the superconductor as a
nonlinear anisotropic conductor\ \cite{Brandt}. In the latter case the
penetration of the magnetic field occurs continuously. In macroscopic
samples the penetration of individual fluxoids is not so important in the
theoretical description of the magnetic response of the superconductor, but
it turns out to be essential in our mesoscopic samples. Recently, Aftalion
and Du\ \cite{Aftalion} studied cylindrical square shaped superconductors
within the Ginzburg-Landau theory. Chibotaru {\it et al} investigated the
vortex entry and the nucleation of anti-vortices in infinite thin
superconducting squares \cite{Chibotaru} and triangles \cite{Chibotaru2}
using the linearized Ginzburg-Landau theory. Within this linear theory they
studied the superconducting/normal transition and they found near this
transition the nucleation of multivortices, anti-vortices and combination of
these two instead of the expected surface superconductivity. They also
calculated the $H-T$ phase diagrams for the square and the triangle.
Recently, Bon\v{c}a and Kabanov \cite{Bonca} studied thin superconducting
squares using the nonlinear Ginzburg-Landau theory in the $\kappa
\rightarrow \infty $ limit. Within this non-linear theory they showed that
the vortex-anti-vortex configuration becomes unstable when moving away from
the superconducting/normal transition.

In the present paper we consider superconductors of finite thickness and
study also the vortex configurations deep inside the superconducting state,
i.e. far from the superconducting/normal boundary. Our main focus will be on
the influence of the geometry of the superconductor on the vortex
configuration and its stability. Our theoretical analysis is based on a full
self-consistent numerical solution of the coupled nonlinear Ginzburg-Landau
equations for arbitrary value of $\kappa $. No a priori shape or arrangement
of the vortex configuration is assumed. The magnetic field profile near and
in the superconductor is obtained self-consistently, and therefore the full
demagnetization effect is included in our approach.

The paper is organized as follows: In Sec.~II we describe the theoretical
formalism. Our results are presented in Sec.~III. We calculate and compare
the free energy and the magnetization for disks, squares and triangles with
the same surface area. Next, we make a distinction between multivortex
states and giant vortex states and we investigate the influence of the
sample geometry on the vortex lattice. We also calculate the magnetic field
range over which the vortex states with vorticity $L$ are stable in disks,
squares and triangles. The magnetic field distribution and the current
density are studied and the $H-T$ phase diagram is obtained. Finally, we
summarized our results in Sec.~IV.

\section{Theoretical formalism}

In the present paper, we consider thin superconducting samples having the
same volume but with different geometry which are immersed in an insulating
medium in the presence of a perpendicular uniform magnetic field $H_{0}$. To
solve this problem we follow the numerical approach of Schweigert and
Peeters \cite{PRL81}. As for thin disks $(d\ll \xi ,\lambda )$ it is allowed
to average the GL equations over the disk thickness for samples of arbitrary
geometry. Using dimensionless variables and the London gauge $div%
\overrightarrow{A}=0$ for the vector potential $\overrightarrow{A}$, we
write the system of GL equations in the following form 
\begin{equation}
\left( -i\overrightarrow{\nabla }_{2D}-\overrightarrow{A}\right) ^{2}\Psi
=\Psi \left( 1-\left| \Psi \right| ^{2}\right) ,  \eqnum{1a}  \label{lijn1}
\end{equation}
\begin{equation}
-\Delta _{3D}\overrightarrow{A}=\frac{d}{\kappa ^{2}}\delta \left( z\right) 
\overrightarrow{j}_{2D},  \eqnum{1b}  \label{lijn2}
\end{equation}
where 
\begin{equation}
\overrightarrow{j}_{2D}=\frac{1}{2i}\left( \Psi ^{\ast }\overrightarrow{%
\nabla }_{2D}\Psi -\Psi \overrightarrow{\nabla }_{2D}\Psi ^{\ast }\right)
-\left| \Psi \right| ^{2}\overrightarrow{A},  \eqnum{1c}  \label{lijn3}
\end{equation}
is the density of superconducting current. The superconducting wavefunction
satisfies the boundary conditions $\left. \left( -i\overrightarrow{\nabla }%
_{2D}-\overrightarrow{A}\right) \Psi \right| _{n}=0$ normal to the sample
surface and $\overrightarrow{A}=\frac{1}{2}H_{0}\rho \overrightarrow{e}%
_{\phi }$ far away from the superconductor. Here the distance is measured in
units of the coherence length $\xi $, the vector potential in $c${\it 
h\hskip-.2em\llap{\protect\rule[1.1ex]{.325em}{.1ex}}\hskip.2em%
}$/2e\xi $, and the magnetic field in $H_{c2}=c${\it 
h\hskip-.2em\llap{\protect\rule[1.1ex]{.325em}{.1ex}}\hskip.2em%
}$/2e\xi ^{2}=\kappa \sqrt{2}H_{c}$. The superconductor is placed in the $%
(x,y)$ plane, the external magnetic field is directed along the $z$ axis,
and the indices 2D, 3D refer to two- and three-dimensional operators,
respectively.

To solve the system of Eqs.~1(a,b), we generalized the approach of Ref.~\cite
{PRL81} for circular disks to superconductors with an arbitrary flat
geometry. We apply a finite-difference representation for the order
parameter and the vector potential on an uniform Cartesian space grid (x,y),
with typically $128\times 128$ grid points for the area of the
superconductor, and use the link variable approach \cite{Kato}, and an
iteration procedure based on the Gauss-Seidel technique to find $\Psi $. The
vector potential is obtained with the fast Fourier transform technique where
we set $\overrightarrow{A}_{\left| x\right| =R_{S},\left| y\right|
=R_{S}}=H_{0}\left( x,-y\right) /2$ at the boundary of a box with a larger
space grid of size typically 4 times the superconductor area.

For circular configurations like disks the giant vortex state is
characterized by the total angular momentum $L$ through $\Psi =\psi \left(
\rho \right) \exp (iL\phi ),$ where $\rho $ and $\phi $ are the cylindrical
coordinates. $L$ is the winding number and gives the vorticity of the
system. Due to the non-linearity of the GL equations an arbitrary
superconducting state is generally a mixture of different angular harmonics $%
L$ even in axially symmetric systems. Nevertheless, we can introduce an
analog to the total angular momentum $L$\ which is still a good quantum
number and which is in fact nothing else then the number of vortices in the
system.

For non-axially symmetric systems there exists no axially symmetric giant
vortex states and hence the superconducting state is always a mixture of
different angular harmonics. The vorticity $L$ of a particular
superconducting sample can be calculated by considering the phase $\varphi $
of the order parameter along a closed loop near the boundary of the sample,
where the total phase difference is always $\Delta \varphi =L\times 2\pi $.
In non-axially symmetric systems three possible vortex states exist: (i) a
multivortex state which contains separate vortices, (ii) a superconducting
state which contains one giant vortex in the center, and (iii) a state which
is a mixture of both: a giant vortex in the center which is surrounded by
single vortices. The giant vortex is not necessary circular symmetric as in
the case of a circular disk, but it may be deformed due to the specific
shape of the sample boundary.

To find the different vortex configurations, which include the metastable
states, we search for the steady-state solutions of Eqs.~1(a,b) starting
from different randomly generated initial conditions. Then we
increase/decrease slowly the magnetic field and recalculate each time the
exact vortex structure. We do this for each vortex configuration in a
magnetic field range where the number of vortices stays the same. By
comparing the dimensionless Gibbs free energies of the different vortex
configurations 
\begin{equation}
F=V^{-1}\int_{V}\left[ 2\left( \overrightarrow{A}-\overrightarrow{A}%
_{0}\right) \cdot \overrightarrow{j}_{2d}-\left| \Psi \right| ^{4}\right] d%
\overrightarrow{r},  \eqnum{2}
\end{equation}
where integration is performed over the sample volume $V,$ and $%
\overrightarrow{A}_{0}$ is the vector potential of the uniform magnetic
field, we find the ground state. The dimensionless magnetization, which is a
direct measure of the expelled magnetic field from the sample, is defined as 
\begin{equation}
M=\frac{\left\langle H\right\rangle -H_{0}}{4\pi },  \eqnum{3}
\label{magnform}
\end{equation}
where $H_{0}$ is the applied magnetic field. $\left\langle H\right\rangle $
is the magnetic field averaged over the sample and $\overrightarrow{H}=rot$ $%
\overrightarrow{A}$.

The temperature is indirectly included in $\xi $, $\lambda $, $H_{c2}$,
whose temperature dependence is given by 
\begin{eqnarray}
\xi (T) &=&\frac{\xi (0)}{\sqrt{\left| 1-T/T_{c0}\right| }}\text{ ,} 
\eqnum{4a} \\
\lambda (T) &=&\frac{\lambda (0)}{\sqrt{\left| 1-T/T_{c0}\right| }}\text{ ,}
\eqnum{4b} \\
H_{c2}(T) &=&H_{c2}(0)\left| 1-\frac{T}{T_{c0}}\right| \text{ ,}  \eqnum{4c}
\end{eqnarray}
where $T_{c0}$ is the critical temperature at zero magnetic field. We will
only explicitly insert temperature if we consider the $H-T$ phase diagrams,
while the other calculations are for a certain fixed temperature. Notice
further that the Ginzburg-Landau parameter $\kappa =\lambda /\xi $ is
independent of the temperature.

\section{Results}

As a typical example, we consider superconducting disks, squares and
triangles with the same surface area $S=\pi 16\xi ^{2}$, the same finite
thickness $d=0.1\xi $ and the same Ginzburg-Landau parameter $\kappa =0.28,$
which is typical for Al \cite{Geim}. Thus the superconducting disk has a
radius $R=4.0\xi $, the square has a width $W=7.090\xi $ and the triangle
has a width $W=10.774\xi $.

\subsection{Free energy and magnetization}

In the first step, we will compare the free energy and the magnetization for
the three geometries. In Figs.~\ref{emag}(a,b) the free energy and the
magnetization are shown for the disk as a function of the applied magnetic
field, in Figs.~\ref{emag}(c,d) for the square and in Figs.~\ref{emag}(e,f)
for the triangle. In Figs.~\ref{emag}(a,c,e) the free energy of the
different giant vortex states is given by solid curves and the free energy
of the multivortex states by dashed curves. The open circles indicate the
transition from multivortex state to giant vortex state at the transition
fields $H_{MG}$. This transition is of second order \cite{PRL81}. Figs.~\ref
{emag}(b,d,f) show the magnetization of the different $L$-states with thin
curves and the ground state is indicated by the thick curve. There exists
vortex states with vorticity up to $L=11$ for the disk and the square and up
to 13 for the triangle. The superconducting state is destroyed at $%
H_{c3}/H_{c2}\approx 1.95$ for the disk, at $H_{c3}/H_{c2}\approx 2.0$ for
the square and $H_{c3}/H_{c2}\approx 2.5$ for the triangle. Thus for samples
with scharp corners the superconducting/normal transition moves to higher
field (for fixed surface area) \cite{PRB60}. Multivortex states can nucleate
in the disk for vorticity $L=2,3,4$ and $5$ and in the square and the
triangle for $L=2,3,4,5$ and $6$. Moreover, for the disk, with increasing
field, the multivortex state always transits to a giant vortex state for
fixed $L$, while for the square and the triangle some $L$-states are
multivortex states over the whole magnetic field range. For the triangle
this is the case for the states with vorticity $L=3,4,5,6$ and for the
square for the states with $L=4$ and $5$. This indicates that breaking the
axial symmetry favors the multivortex state over the giant vortex state. In
some magnetic field regions, the vortex states exhibitis a paramagnetic
response, i.e. $-M<0$. This occurs in the disk for metastable states with $%
L=1$, $4-8$ and in the square for metastable states with $L=1$, $4$, $5$.
For the triangle $-M$ is always positive, i.e. only diamagnetic behaviour is
observed.

In Fig.~\ref{flux}(a) we plot the magnetic flux $\phi _{L\rightarrow L+1}$
passing through the superconductor at the thermodynamic transition field $%
H_{L\rightarrow L+1}$ as a function of the vorticity $L$ for the three
geometries. We scaled this flux by $(L+1)\phi _{0}$ which is the expected
flux when no boundary effects are important. The result for the disk is
given by circles, for the square by squares and for the triangle by
triangles. In the figure we use open symbols for transitions between two
giant vortex states, filled symbols for transitions between a multivortex
and a giant vortex state, and crossed symbols for transitions between two
multivortices. The symbol for the largest $L$ corresponds to the
superconducting/normal transition. For the disk we find $L\rightarrow L+1$
transitions between a giant vortex and a multivortex state for $L=1$ and $2$%
, while the other transitions occur between two giant vortex states. For the
square we find transitions between a multivortex state and a giant vortex
state for $L=2$, $3$, $6$ and between two multivortex states for $L=4$ and $5
$. For the triangle, transitions between a giant and a multivortex state
occur for $L=1$, $2$ and $6$ and between two multivortex states for $L=2$, $3
$ and $4$. Notice further that for $L>6$, the value of $\phi _{L\rightarrow
L+1}/(L+1)\phi _{0}$ becomes almost independent of $L$ for the three
geometries: $\phi _{L\rightarrow L+1}/(L+1)\phi _{0}\approx 1.3$ for the
disk and the square and $\phi _{L\rightarrow L+1}/(L+1)\phi _{0}\approx 1.4$
for the triangle. The magnetic flux $\phi _{L\rightarrow L+1}$ passing
through the superconductor at the thermodynamic transition field $%
H_{L\rightarrow L+1}$ can be fitted for the three geometries by the
following formula 
\[
\frac{\phi _{L\rightarrow L+1}}{(L+1)\phi _{0}}=\frac{a+bL}{1+cL}\text{ ,}
\]
with $a=3.10837,$ $b=0.61792$ and $c=0.59825$ for the disk (solid curve in
Fig.~\ref{flux}(a)), $a=3.19769,$ $b=0.60962$ and $c=0.58396$ for the square
(dashed curve), and $a=3.58614,$ $b=0.66114$ and $c=0.6031$ for the triangle
(dotted curve).

The amount of flux increase needed to increase the vorticity of the ground
state from $L$ to $L+1$ is plotted in Fig.~\ref{flux}(b) for the disk (solid
curves, open circles), the square (dashed curve, open squares) and the
triangle (dotted curve, open triangles). For $3<L<11$, this value is almost
independent of $L$ and of the geometry and equals $\Delta \phi /\phi
_{0}\simeq 1.1$, i.e. it is approximately equal to, but still different
from, the flux quantum $\phi _{0}$. Notice that $\Delta \phi /\phi _{0}>1$
for any $L$ which is a consequence of the demagnetization effects.

\subsection{Multivortex states}

To distinguish whether the superconducting state is a multivortex state or a
giant vortex state and to determine the multivortex to giant vortex state
transition field, we considered the Cooper-pair density $\left| \Psi \right|
_{center}^{2}$ in the center of the superconductor \cite{PRL81}. We can be
sure that the superconducting state is a multivortex state if $\left| \Psi
\right| _{center}^{2}\neq 0$ for $L>1$. The reason is that giant vortices
are always in the center of the superconductor, and hence $\left| \Psi
\right| _{center}^{2}=0$ for giant vortex states. Fig.~\ref{denscent} shows
the Cooper-pair density in the center of the square as a function of the
applied magnetic field. The Cooper-pair density $\left| \Psi \right|
_{center}^{2}$ is finite for $H_{0}/H_{c2}<H_{GM}/H_{c2}\approx 0.5825$ and $%
0.7825$ for $L=2$ and $3$, respectively, and $\left| \Psi \right|
_{center}^{2}=0$ for $H_{0}/H_{c2}>H_{GM}/H_{c2}$. For $L=4$ the Cooper-pair
density in the center differs from zero over the whole magnetic field region
where the $L=4$ state is stable. On the other hand $\left| \Psi \right|
_{center}^{2}=0$ does not guarantee that the superconducting state is a
giant vortex state. For example, the multivortex state with $L=5$ in a
square shows 4 vortices away from the center situated on the diagonals and
one vortex in the center, and hence $\left| \Psi \right| _{center}^{2}=0$.
Therefore, we studied the Cooper-pair density distribution in detail. If two
vortices are very close to each other, then the Cooper-pair density on the
axis between these two vortices will become very low too, which means that
the separation between two vortices becomes invisible in the contourplots.
Therefore we have to define another criterion to determine the multivortex
to giant vortex transition. If the maximum between two minima in the
Cooper-pair density (i.e. the vortices) is lower than $0.5\%$ of the maximum
Cooper-pair density $\left| \Psi \right| _{max}^{2}$ in the sample, then we
will say that the vortices form a giant vortex state instead of a
multivortex state. With this criterion we find that the $L=5$ state in a
square is always a multivortex state and the $L=6$ state is a multivortex
state for $H_{0}/H_{c2}<1.37$ and a giant vortex state for $%
H_{0}/H_{c2}>1.37 $.

How do the multivortex states look like? Figs.~\ref{contv}(a-c) show the
Cooper-pair density for a multivortex state with $L=2$, $3$, and $4$ at $%
H_{0}/H_{c2}=0.42$, $0.67$, and $0.745,$ respectively. High Cooper-pair
density is given by dark regions, low Cooper-pair density by light regions.
For $L=2$ the vortices are along the diagonal, for $L=3$ the vortices are on
a triangle, and for $L=4$ they are on a square. For $L=4$ only the
multivortex state is found which is favoured over the giant vortex state.
The reason is that the square vortex lattice easily fits in the sample.
Figs.~\ref{contv}(d,e) show the phase of the order parameter for the
multivortex states with $L=5$ at $H_{0}/H_{c2}=0.82$ and with $L=6$ at $%
H_{0}/H_{c2}=1.32$. Phases near zero are given by light regions and phases
near $2\pi $ by dark regions. By going around the superconductor, the phase
cleary changes 5 times $2\pi $ in Fig.~\ref{contv}(d) and 6 times $2\pi $ in
Fig.~\ref{contv}(e). For $L=5$ there are 4 vortices on a square and the
fifth vortex is in the center. The latter has clearly vorticity one. For $L=6
$ there are also 4 vortices on a square and the other vortices are in the
center forming one giant vortex with vorticity 2. Thus in this case we have
the remarkable coexistence of a giant vortex in the center with vorticity 2
and 4 clearly separated vortices around it. For this case the multivortex to
giant vortex transition field is defined as the field where separate
vortices appear with decreasing field. This means that in Fig.~\ref{emag}
some states are indicated as multivortex, even though there exist a (giant)
vortex in the center with vorticity $L>1$. Thus we consider a state no
longer as a giant vortex state when not all flux of the vortex is confined
in a single connected region. Notice also that not only the configuration of
the multivortices tries to have the same geometry as the sample, but also
the giant vortex geometry depends on the sample geometry.

We found that for this size of the square sample, the states with $L>6$ are
always in the giant vortex state. With increasing $L$ this giant vortex
grows and superconductivity only occurs in the corners of the square. This
is illustrated in Figs.~\ref{contv11}(a,b) which shows the Cooper-pair
density for the $L=11$ state at $H_{0}/H_{c2}\approx 1.9$ and $1.95$. It is
obvious that further increasing the field pushes the superconducting
condensate more to the corners. At the superconducting/normal transition
field $H_{0}/H_{c2}\approx 2.0$ the corners becomes normal too.

Fig.~\ref{squareL3} shows the positions of the vortices for the $L=3$ state
in a square at applied magnetic fields $H_{0}/H_{c2}=0.545$, $0.62$, $0.695$%
, and $0.77$. The latter one is just below the multivortex to giant vortex
state transition field $H_{GM}/H_{c2}\approx 0.7825$. The solid lines
indicate the square boundaries. With increasing field the vortices move
towards the center of the square and at $H_{GM}/H_{c2}\approx 0.7825$ they
combine in the center and form one giant vortex with vorticity $L=3.$ In the
multivortex state, one vortex is always situated on the diagonal of the
square, regardless of the magnetic field. The other two vortices are located
such that the three vortices form a equilateral triangle which is centered
in the center of the square. Since the vortices move to the center with
increasing field, the width $W$ of the triangular vortex lattice decreases,
i.e. $W=3.27,$ $2.89$, $2.32$, $1.61$ at $H_{0}/H_{c2}=0.545$, $0.62$, $%
0.695 $, $0.77$, respectively.

For the triangle geometry multivortex states nucleate with vorticity $%
L=2,3,4,5$ and $6$ (see Figs.~\ref{emag}(e,f)). Figs.~\ref{contd}(a-c) show
the Cooper-pair density for a multivortex state with $L=2$, $3$, and $4$ at $%
H_{0}/H_{c2}=0.495$, $0.82$, and $0.745,$ respectively. High Cooper-pair
density is given by dark regions and low Cooper-pair density by light
regions. In the multivortex state with vorticity $L=2$ the vortices are
situated along one of the perpendicular bisectors of the triangle. In the $%
L=3$ state the vortices are on a triangle which easily fits in the sample,
while the $L=4$ state consists of 3 vortices on a triangle and the fourth
vortex is situated in the center. Instead of the square configuration as in
the case of the square geometry, the vortex lattice tries to copy the
geometry of the sample, i.e. the triangular geometry. For the multivortex
states with $L=5$ and $L=6$, the separation of vortices becomes invisible in
the contourplots of the Cooper-pair density, which show one big vortex in
the center. The reason is that the maximum Cooper-pair density on the axis
between two vortices is very low. Therefore, we show the phase of the order
parameter in Figs.~\ref{contd}(d,e) for the multivortex states with $L=5$ at 
$H_{0}/H_{c2}=1.27$ and with $L=6$ at $H_{0}/H_{c2}=1.345$. Phases near zero
are given by light regions and phases near $2\pi $ by dark regions. In both
cases there is a coexistence of a giant vortex in the center and 3 separated
multivortices around it placed in the direction of the corners. Taking a
loop around the giant vortex, the phase changes respectively by 2 and 3
times $2\pi $, which means that the vorticity of the giant vortex is 2 in
the case of the $L=5$ multivortex state and 3 in the case of the $L=6$
multivortex state. Notice also that for $L=6$ the geometry of the giant
vortex is not axial symmetrical, but triangular. States with $L>6$ are
always giant vortex states as we also found for the square. With increasing $%
L$ this giant vortex grows and for large vorticities superconductivity only
occurs in the corners of the triangle. Further increasing the field pushes
the superconductivity more to the corner untill these corners becomes normal
too at the superconducting/normal transition field.

For the circular disk we find multivortex states with vorticity $L=2,3,4$
and $5$ (see Figs.~\ref{emag}(a,b)). Figs.~\ref{contc}(a-d) show the
Cooper-pair density for the multivortex states with vorticity $L=2,3,4,5$ at 
$H_{0}/H_{c2}=0.495$, $0.62$, $0.965$ and $0.82$, respectively. High
Cooper-pair density is given by dark regions, low by light regions.
Multivortex states for disks were already studied in previous papers (see
introduction). Therefore, in this paper we only stress that in a disk the
multivortices are positioned on a ring, which means that also in this case
the sample imposes its symmetry on the vortex lattice.

From the study of the Cooper-pair density and the phase of the order
parameter we learned that: (i) multivortex states nucleate in disks as well
as in squares and triangles for several values of the vorticity $L$, and
(ii) the vortex lattices tries to have the same geometry as the sample.

\subsection{Magnetic field distribution - Demagnetization effects}

Since we studied samples with finite thicknesses, demagnetization effects
are important and therefore we had to solve for the magnetic field
distribution around the sample. In this section we will describe the
magnetic field distribution for the square. The results for the disk and the
triangle are analogous.

In Figs.~\ref{magv23}(a-f) the magnetic field distribution is shown for the
square geometry for the state with vorticity $L=2$ at $H_{0}/H_{c2}=0.42$, $%
0.52$ and $0.62$ (see open circles in Fig.~\ref{denscent}), and with
vorticity $L=3$ at $H_{0}/H_{c2}=0.62$, $0.72$ and $0.82$, respectively.
High magnetic field is given by dark regions and low by light regions. The
magnetic field is clearly non uniform in and around the sample. The dark
spots in the square are the vortices and the dark regions near the sample
surface are due to the compression of the magnetic lines when they are
forced to go around the sample. These regions are responsible for the
demagnetization effects. It is clear that with increasing external field and
fixed number of vortices, the demagnetization effects are more pronounced,
because the superconductor has to expel more magnetic field. In Figs.~\ref
{magv23}(a,b,d,e) the superconducting state is a multivortex state and the
separated vortices are clearly visible, while in Figs.~\ref{magv23}(c,f)
where $H_{0}/H_{c2}>H_{MG}/H_{c2}\approx 0.5825$ and $0.7825$ for $L=2$ and $%
3$, respectively, there is one giant vortex in the center. With increasing
field the vortices move towards the center and at $H_{0}=H_{MG}$ they
combine to one giant vortex state. Notice that the giant vortex state is not
necessarily axial symmetric as was in the case of the disk.

In Figs.~\ref{magv4}(a-d) the magnetic field distribution is shown for the
square geometry for the state with vorticity $L=4$ at the magnetic fields
indicated by the open circles in Fig.~\ref{denscent}, i.e. $%
H_{0}/H_{c2}=0.72 $, $0.82$, $0.92$ and $1.02$, respectively. Now, there is
no transition from multivortex to giant vortex state and the four vortices
are cleary visible as the dark spots. Notice that from the magnetic field
distribution one clearly observes that the vortex lattice is a square
lattice, i.e. the lattice geometry is the same as the sample geometry, and
that the vortices move towards the center with increasing field.

For the multivortex states with higher vorticity, the separated
multivortices are not visible anymore in the contour plots of the magnetic
field distribution. The problem is the same as for the contour plots of the
Cooper-pair density, i.e. the vortices are too close to each other and the
spots corresponding to high magnetic fields are overlapping in the picture.
For high vorticity and high external fields, the total magnetic field
appreciably differs from the externally applied field only in the corners of
the square. Figs.~\ref{magv11}(a,b) shows the magnetic field distribution
for the same configuration as in Figs.~\ref{contv11}(a,b), i.e. the $L=11$
state at $H_{0}/H_{c2}\approx 1.9$ and $1.95$, respectively. A local
decrease in magnetic field is given by the light regions, an increase by the
dark regions. In both pictures the magnetic field is only substantially
expelled in the corners and consequently only near the corners there is a
higher density of magnetic field lines at the outside of the square. Further
increasing the field destroys the superconductivity and, thus, the total
field becomes equal to the external one over the whole sample.

Next, we investigate the dependence of the magnetic field on $z$. Figs.~\ref
{magneetz}(a-f) show the magnetic field distribution for the $L=4$ state in
a square for different values of $z,$ $z/\xi =0.0$, $0.1$, $0.3$, $0.6$, $%
1.0 $ and $10.0$, respectively. The applied magnetic field is $%
H_{0}/H_{c2}=0.77$. High magnetic field is given by dark regions, low
magnetic field by light regions. In the plane of the superconductor, i.e. $%
z=0$, the magnetic field which penetrates the superconductor is either
compressed into multivortices or expelled to the outside of the sample.
Therefore, the four dark spots in Fig.~\ref{magneetz}(a) indicate the
vortices and the light regions towards the sample boundary are due to the
expulsion of the magnetic field towards the outside of the superconductor.
As a consequence, the magnetic field increases in a small strip near the
sample boundary. With increasing $z$ and $\left| z\right| >d/2$, the
magnetic field will still be influenced by the superconductor. The
demagnetization effects decrease with increasing $z$ and the compression of
the magnetic field lines into vortices becomes smaller. Therefore, the
vortices and the expulsion of the field will become less pronounced with
increasing $z$. At $z=0.1\xi $, the vortices and the results of the magnetic
field expulsion are still visible by the dark and light regions (see Fig.~%
\ref{magneetz}(b)). In Fig.~\ref{magneetz}(c) and (d), at $z=0.3\xi $ and $%
0.6\xi $, respectively, the contrast in the picture decreases which means
that the influence of the superconductor, i.e. the compression and expulsion
of the magnetic field lines, decreases. At $z=1.0\xi $ the magnetic field
just slightly decreases right above the superconductor compared to the
external field (see Fig.~\ref{magneetz}(e)). At $z=10.0\xi $ the total
magnetic field is homogeneous. Far away from the superconductor, the
magnetic field is not influenced by the superconductor and equals the
external field. This is clearly shown in Fig.~\ref{magneetz}(f).

\subsection{Superconducting current density}

When a superconducting sample is placed in an external magnetic field, the
magnetic field is expelled from the superconductor due to screening currents
near the sample boundary. The direction of the screening currents is such
that the corresponding magnetic field is opposite to the external one, which
leads to a lower total field in the superconductor. Magnetic field
penetrating the superconductor creates currents flowing in the other
direction than the screening currents. The competition between these
currents and the screening currents results in the existence of vortices.

Figs.~\ref{currv}(a-d) show vector plots of the supercurrent in the
superconducting square for the $L=1$ state at $H_{0}/H_{c2}=0.27$, the $L=2$
state at $H_{0}/H_{c2}=0.42$, the $L=3$ state at $H_{0}/H_{c2}=0.67$ and the 
$L=4$ state at $H_{0}/H_{c2}=0.745$, respectively. Figs.~\ref{currv}(e-h)
show the corresponding contour plots of the phase of the order parameter.
Phases near $2\pi $ are given by dark regions, phases near zero by light
regions. From the phase of the order parameter one can easily determine the
number and the positions of the vortices. In Fig.~\ref{currv}(a) it is clear
that the screening currents near the sample boundary flow clockwise and the
currents around the vortex in the center counterclockwise. In Figs.~\ref
{currv}(b,c,d) there are currents flowing counterclockwise around $2,$ $3$
and $4$ vortices, respectively. Around one vortex, the size of the current,
indicated by the length of the arrows in Figs.~\ref{currv}(a-d), is not the
same for every angle. In Fig.~\ref{currv}(b) it is clear that in the region
between the two vortices the currents around these two vortices cancel each
other out. Also, in the case of $L=3$ and $L=4$ the currents around the
different vortices cancel out each other in the center of the sample (see
Fig.~\ref{currv}(c,d)).

From Figs.~\ref{currv}(a-d) one expects anti-vortices towards the corners,
because there are some spots where the currents flow in clockwise direction.
That these are not really anti-vortices can be seen from the phase of the
order parameter (Figs.~\ref{currv}(e-h)). By going around an anti-vortex,
the phase changes with $-2\pi $ and this is clearly not the case here.
Moreover, the Cooper-pair density, shown in Figs.~\ref{contv}(a-c), is not
zero at these positions.

Next, we investigate the superconducting current density in the triangular
sample. Figs.~\ref{currd}(a-c) show vector plots of the supercurrent in the
superconducting triangle for the $L=1$ state at $H_{0}/H_{c2}=0.27$, the $%
L=2 $ state at $H_{0}/H_{c2}=0.495$, and the $L=3$ state at $%
H_{0}/H_{c2}=0.82$, respectively. Figs.~\ref{currd}(d-f) show the contour
plots of the corresponding phase of the order parameter. Phases near $2\pi $
are given by dark regions, phases near zero by light regions. The behavior
of the supercurrent in triangular samples is similar to the one in square
samples. The screening currents flow clockwise and the current around the
vortices in the opposite direction. The currents around different vortices
cancel each other in the region between them. Towards the corners, there are
some spots where the current flows in a clockwise direction, but these spots
are not anti-vortices. This can be seen from the phase of the order
parameter (see Figs.~\ref{currd}(d-f)) and from the Cooper-pair density (see
Figs.~\ref{contd}(a-b)).

\subsection{Stability of the vortex states}

Not only the stability region of the multivortex states with respect to the
giant vortex states depends on the sample geometry, but also the stability
of each individual superconducting state is sensitive to the geometry. In
Fig.~\ref{stab} we show the magnetic field range $\Delta H$ over which the
vortex state with vorticity $L$ is stable, i.e. $\Delta
H=H_{expulsion}-H_{penetration}$ (see also Figs.~\ref{pen_exp}(a,b)), as a
function of the vorticity $L$, for $L\leq 6$ and in the inset for $L\geq 6$.
For the disk the result is shown by the open circles, for the square by the
open squares and for the triangle by the open triangles where the curves are
guides to the eye. For the circular disk the stability region $\Delta
H/H_{c2}$ uniformly decreases with increasing $L$ with a slight dip at $%
L=2,3 $. The square and the triangle exhibit a peak structure in the region $%
L<5$. For the square we find that the state with $L=4$ is stable over a
larger magnetic field region than the state with vorticity $L=3$, which is a
consequence of the fact that the vortex lattice tries to keep the same
geometry as the sample. For the triangle we find a peak at $L=3$ and a dip
at $L=2$ for the same reason. Notice that: i) the peak structure is more
pronounced for structures which fit more closely the triangular Abrikosov
lattice; ii) for $L>4$ no clear peaks are found; iii) the vortex states in
the square and circle geometry have almost the same stability range for $%
L\leq 2$ and $L\geq 6$; iv) for $L\geq 4$ the stability range for the vortex
state in the triangular geometry becomes substantially smaller than for the
other two geometries which have less sharp corners. Thus sharp corners
decrease the stability range of the vortex states, i.e. it is easier for
flux to enter the system; and v) Fig.~\ref{flux}, which shows the extra flux
needed to increase the vorticity by one unit, contains complementary
information to Fig.~\ref{stab}.

\subsection{H-T phase diagram}

Untill now, all our calculations were done for fixed temperature $T$. Now we
will include temperature and our lateral dimensions and fields will be
expressed in the zero temperature results $\xi (0)$ and $H_{c2}(0)$,
respectively. Temperature will be expressed in units of the zero magnetic
field criticial temperature $T_{c0}$. We take the surface area of our
samples $S=16\pi \xi ^{2}(0)$ and the thickness $d=0.1\xi (0)$.

The $H-T$ phase diagram is shown in the inset of Fig.~\ref{htfase} for the
disk (solid curves), the square (dashed curves) and the triangle
(dash-dotted curves) for the states with vorticity $L=0$ and $L=1$, thus for
low fields and temperatures close to $T_{c0}$. The thick curves are the
superconducting/normal transitions and the thinner curves indicate the
expulsion and the penetration fields, i.e. the boundaries of the stabilitiy
region of the state with vorticity $L$. The lower thin curves show the
transition from the state with vorticity $L=1$ to $L=0$ with decreasing
field (expulsion) and the upper thin curves show the transition from $L=0$
to $L=1$ with increasing field (penetration). Fig.~\ref{htfase} shows the $%
H-T$ phase diagram for higher fields. For the sake of clarity only the
superconducting/normal transition $H_{c3}$ is shown as a function of
temperature. The black dots indicate the transitions fields between the
different $L$-states. For every (fixed) temperature the
superconducting/normal transition field is highest for the triangle and
lowest for the disk. For every (fixed) magnetic field, the critical
temperature is highest for the triangle and lowest for the disk. This means
that with sharper corners, the critical temperature and critical field are
enhanced due to an enhanced surface superconductivity. These results are in
good agreement with the phase diagrams found in Ref.~\cite
{Bruyndoncx2,Chibotaru,Chibotaru2}.

\section{Conclusions}

We investigated theoretically the influences of the geometry of thin
superconducting samples on the vortex configuration. Therefore, we
considered superconducting disks, squares and triangles with the same
surface area $S=\pi 16\xi ^{2}$ and the same thickness $d=0.1\xi $ for $%
\kappa =0.28$. For these three geometries we calculated the free energy and
the magnetization of the different giant and multivortex states as a
function of the applied magnetic field and we indicated the multivortex to
giant vortex transitions for fixed vorticity $L$. Multivortex states were
found for disks as well as for squares and triangles for several values of
the vorticity. For given $L$, the vortex lattice was different in the three
geometries due to the fact that it tries to adapt to the geometry of the
sample. This influences considerably the stability range of the different
vortex states. For squares and triangles we found magnetic field regions
where there is a coexistence between a giant vortex state in the center and
several separated vortices in the direction of the sample corners. Near the
superconducting/normal transition we do not find multivortices,
anti-vortices or a combination of them, but we find surface
superconductivity. We studied the magnetic field distribution across the
superconductor and around the superconductor which clearly shows the
demagnetization effects, which are very important for samples of finite
thickness. The vector plots of the superconducting current showed spots
where the current flows in clockwise direction. From the phase of the order
parameter and the Cooper-pair density we conclude that these spots are not
anti-vortices. We also investigated the stability of the vortex states with
vorticity $L$ by calculating the magnetic field range over which the vortex
states with vorticity $L$ are stable. We found that this stability range
sensitively depends on the sample geometry. As a function of $L$ we found
enhanced stability for the triangle for $L=3$ and for the square for $L=4$.
In the last section, we also included temperature by calculating a $H-T$
phase diagram for the disk, the square and the triangle. With sharper sample
corners, we found that for fixed temperature, the superconducting/normal
transition field $H_{c3}$ moves to higher fields, and for fixed field, the
critical temperature increases.

\section{Acknowledgments}

This work was supported by the Flemish Science Foundation (FWO-Vl), the
''Onderzoeksraad van de Universiteit Antwerpen'' (GOA), the
''Interuniversity Poles of Attraction Program - Belgian State, Prime
Minister's Office - Federal Office for Scientific, Technical and Cultural
Affairs'', and the European ESF-Vortex Matter. Discussions with S.
Yampolskii are gratefully acknowledged.

\bigskip

\begin{figure}[tbp]
\caption{The free energy and the magnetization as a function of the applied
magnetic field for the disk, the square, and the triangle with the same
surface area $S=\protect\pi 16\protect\xi ^{2}$ and thickness $d=0.1\protect%
\xi $ for $\protect\kappa =0.28$. (a,c,e) The free energy of the giant
vortex state (solid curves) and the multivortex states (dashed curves) and
the multivortex to giant vortex transition fields (open circles). (b,d,f)
The magnetization of the $L$-states (solid curves) and the ground state
(thick solid curve).}
\label{emag}
\end{figure}

\begin{figure}[tbp]
\caption{(a) The flux $\protect\phi _{L\rightarrow L+1}$ corresponding to
the thermodynamic transition field $H_{L\rightarrow L+1}$ as a function of
the vorticity $L$ for the disk (circles), for the square (squares) and for
the triangle (triangles). The curves are the fitted results. (b) The flux $%
\Delta \protect\phi $ corresponding to the magnetic field range needed to
increase the vorticity of the ground state from $L$ to $L+1$ as a function
of the vorticity $L$ for the disk (solid curves, open circles), the square
(dashed curve, open squares) and the triangle (dotted curve, open
triangles). }
\label{flux}
\end{figure}

\begin{figure}[tbp]
\caption{The Cooper-pair density in the center of the square as a function
of the applied magnetic field for the states with vorticity $L=2,3$ and $4$.}
\label{denscent}
\end{figure}

\begin{figure}[tbp]
\caption{(a-c) The Cooper-pair density for a multivortex state in a square
with $L=2$, $3$, and $4$ at $H_{0}/H_{c2}=0.42$, $0.67$, and $0.745,$
respectively. High Cooper-pair density is given by dark regions, low
Cooper-pair density by light regions. (d,e) The phase of the order parameter
for the multivortex states with $L=5$ at $H_{0}/H_{c2}=0.82$ and with $L=6$
at $H_{0}/H_{c2}=1.32$. Phases near zero are given by light regions, phases
near $2\protect\pi $ by dark regions.}
\label{contv}
\end{figure}

\begin{figure}[tbp]
\caption{The Cooper-pair density for the $L=11$ state in a square at $%
H_{0}/H_{c2}\approx 1.9$ (a) and $1.95$ (b). }
\label{contv11}
\end{figure}

\begin{figure}[tbp]
\caption{The multivortex positions of the $L=3$ state in a square at applied
magnetic fields $H_{0}/H_{c2}=0.545$, $0.62$, $0.695$, and $0.77.$}
\label{squareL3}
\end{figure}

\begin{figure}[tbp]
\caption{(a-c) The Cooper-pair density for the multivortex states in a
triangle with $L=2$, $3$, and $4$ at $H_{0}/H_{c2}=0.495$, $0.82$, and $%
0.745,$ respectively. High Cooper-pair density is given by dark regions and
low Cooper-pair density by light regions. (d,e) The phase of the order
parameter for the multivortex states with $L=5$ at $H_{0}/H_{c2}=1.27$ and
with $L=6$ at $H_{0}/H_{c2}=1.345$. Phases near zero are given by light
regions, phase near $2\protect\pi $ by dark regions. }
\label{contd}
\end{figure}

\begin{figure}[tbp]
\caption{(a-d) The Cooper-pair density for the multivortex states in a disk
with vorticity $L=2,3,4,5$ at $H_{0}/H_{c2}=0.495$, $0.62$, $0.965$ and $%
0.82 $, respectively. High Cooper-pair density is given by dark regions, low
by light regions. }
\label{contc}
\end{figure}

\begin{figure}[tbp]
\caption{The magnetic field distribution for the square for the state with
vorticity $L=2$ at $H_{0}/H_{c2}=0.42$ (a), $0.52$ (b) and $0.62$ (c), and
with vorticity $L=3$ at $H_{0}/H_{c2}=0.62$ (d), $0.72$ (e) and $0.82$ (f),
respectively. High magnetic field is given by dark regions and low by light
regions.}
\label{magv23}
\end{figure}

\begin{figure}[tbp]
\caption{The magnetic field distribution for the square for the state with
vorticity $L=4$ at $H_{0}/H_{c2}=0.72$ (a), $0.82$ (b), $0.92$ (c) and $1.02$
(d). High (low) magnetic fields are given by dark (light) regions.}
\label{magv4}
\end{figure}

\begin{figure}[tbp]
\caption{The magnetic field distribution for the $L=11$ state in a square at 
$H_{0}/H_{c2}\approx 1.9$ (a) and $1.95$ (b). High (low) magnetic fields are
given by dark (light) regions.}
\label{magv11}
\end{figure}

\begin{figure}[tbp]
\caption{The magnetic field distribution for the $L=4$ state in a square for
different values of $z$; $z/\protect\xi =0.0$ (a), $0.1$ (b), $0.3$ (c), $%
0.6 $ (d), $1.0$ (e) and $10.0$ (f), respectively. The applied magnetic
field is $H_{0}/H_{c2}=0.77$. High magnetic field is given by dark regions,
low magnetic field by light regions.}
\label{magneetz}
\end{figure}

\begin{figure}[tbp]
\caption{(a-d) Vector plots of the supercurrent in the superconducting
square and\ (e-h) the contour plots of the phase of the order parameter for
the $L=1$ state at $H_{0}/H_{c2}=0.27$ (a,e), the $L=2$ state at $%
H_{0}/H_{c2}=0.42$ (b,f), the $L=3$ state at $H_{0}/H_{c2}=0.67$ (c,g) and
the $L=4$ state at $H_{0}/H_{c2}=0.745$ (d,h). Phases near $2\protect\pi $
are given by dark regions, phases near zero by light regions.}
\label{currv}
\end{figure}

\begin{figure}[tbp]
\caption{(a-c) Vector plots of the supercurrent in the superconducting
triangle and\ (d-f) the contour plots of the phase of the order parameter
for the $L=1$ state at $H_{0}/H_{c2}=0.27$ (a,d), the $L=2$ state at $%
H_{0}/H_{c2}=0.495$ (b,e), and the $L=3$ state at $H_{0}/H_{c2}=0.82$ (c,f).
Phases near $2\protect\pi $ are given by dark regions, phases near zero by
light regions.}
\label{currd}
\end{figure}

\begin{figure}[tbp]
\caption{The magnetic field range over which the vortex states with
vorticity $L$ are stable as a function of the vorticity $L$ for the disk
(open circles, solid curves), the square (open squares, dashed curves) and
the triangle (open triangles, dotted curves).}
\label{stab}
\end{figure}

\begin{figure}[tbp]
\caption{(a) The expulsion field and (b) the penetration field as a function
of the vorticity $L$ for disk (solid curves, circles), the square (dashed
curves, squares) and the triangle (dotted curves, triangles).}
\label{pen_exp}
\end{figure}

\begin{figure}[tbp]
\caption{The $H-T$ phase diagram for the disk (solid curve), the square
(dashed curve) and the triangle (dash-dotted curve). Only the
superconducting/normal transition $H_{c3}$ is shown as a function of
temperature. The black dots indicate the transitions. The inset shows the $%
H-T$ phase diagram for the states with vorticity $L=0$ and $L=1$. The thick
curves are the superconducting/normal transitions and the thinner curves
indicate the expulsion and the penetration.}
\label{htfase}
\end{figure}

\end{document}